\documentclass[conference]{IEEEtran}
\IEEEoverridecommandlockouts
\usepackage{cite}
\usepackage{amsmath,amssymb,amsfonts}
\usepackage{algorithmic}
\usepackage{graphicx}
\usepackage{textcomp}
\usepackage{xcolor}

\usepackage{braket}
\usepackage{quantikz}
\usepackage{physics}
\usepackage{url}

\usepackage{tabularx} 
\usepackage{booktabs}
\usepackage{makecell}
\usepackage{xcolor}
\usepackage{balance}

\def\BibTeX{{\rm B\kern-.05em{\sc i\kern-.025em b}\kern-.08em
    T\kern-.1667em\lower.7ex\hbox{E}\kern-.125emX}}
\begin{document}

\title{Digital Annealer-Assisted CNOT-Count-Oriented \\
Quantum Circuit Transpilation \\
with Integrated QUBO Mapping and Routing}


\author{\IEEEauthorblockN{Kazuma Watanabe}
\IEEEauthorblockA{
\textit{Keio University}\\
Yokohama, Japan \\
k.watanabe0709@keio.jp}
\and
\IEEEauthorblockN{Hideaki Kawaguchi}
\IEEEauthorblockA{
\textit{Keio University}\\
Yokohama, Japan \\
hikawaguchi@keio.jp}
\and
\IEEEauthorblockN{Shin Nishio}
\IEEEauthorblockA{
\textit{Keio University}\\
Yokohama, Japan  and\\
\textit{University College London}\\
London, UK\\
shin.nishio@keio.jp}
\and
\IEEEauthorblockN{Takahiko Satoh}
\IEEEauthorblockA{
\textit{Keio University}\\
Yokohama, Japan \\
satoh@keio.jp}
}

\maketitle

\begin{center}
\footnotesize
© 2026 IEEE. Personal use of this material is permitted.
Permission from IEEE must be obtained for all other uses, in any current or
future media, including reprinting/republishing this material for advertising
or promotional purposes, creating new collective works, for resale or
redistribution to servers or lists, or reuse of any copyrighted component of
this work in other works.
\end{center}

\begin{abstract}
Limited qubit counts and two-qubit gate errors motivate reducing CNOT overhead in noisy intermediate-scale quantum (NISQ) transpilation. In circuit-partitioning workflows, repeated hardware sampling can motivate additional classical search, although this study does not evaluate end-to-end runtime or hardware fidelity. We evaluate two Digital Annealer (DA) strategies on seven fixed standalone benchmark instances and a 64-node $8\times8$ square-grid. Hybrid combines DA-based global initial mapping with Qiskit routing, whereas Full DA applies DA to both initial mapping and iterative routing.

Hybrid achieved lower mean equivalent CNOT counts than both configured baselines on four benchmarks and identical means on the remaining three. The unweighted macro-average reductions were 6.14\% relative to Qiskit and 17.9\% relative to t$\vert$ket$\rangle$. Among the six benchmarks with feasible Full DA outputs, Full DA achieved lower means than ISAAQ on five, with a success-conditioned macro-average reduction of 48.9\%; however, it produced no feasible BV output and did not outperform Hybrid on any benchmark. These results identify DA-generated global placement followed by heuristic routing as the primary positive result and characterize the limitations of the evaluated short-horizon DA routing workflow. Equivalent CNOT count is a structural proxy and does not directly measure execution fidelity.
\end{abstract}
\section{Introduction}
\label{sec:intro}

Limited qubit counts and high two-qubit gate error rates constrain the scalability of noisy intermediate-scale quantum (NISQ) devices~\cite{Preskill2018quantumcomputingin}. Circuit partitioning enables larger circuits to execute by decomposing them into smaller subcircuits~\cite{Peng_2020, Mitarai_2021, Mitarai2021overheadsimulating}. However, reconstructing expectation values through quasi-probability sampling introduces an overhead that scales approximately as \(9^{M_s}16^{M_t}\) for \(M_s\) gate cuts and \(M_t\) wire cuts~\cite{Mitarai_2021, Mitarai2021overheadsimulating}. Repeated hardware executions can therefore dominate the one-time classical compilation cost, motivating additional classical search to reduce two-qubit-gate overhead. Accordingly, we evaluate equivalent-CNOT reduction rather than compilation time. Equivalent CNOT count is used only as a structural proxy; reconstructed accuracy, hardware fidelity, and the end-to-end runtime tradeoff are outside the scope of this paper.

Mapping and routing are NP-hard~\cite{Tillmann_2016, botea2018complexity}. Qiskit SABRE~\cite{qiskit231, qiskit2024, li2019sabre} and t$\vert$ket$\rangle$~\cite{Sivarajah_2020} use finite heuristic search budgets, so allocating additional classical computation to placement may ultimately yield a lower gate count.

We evaluate Fujitsu's Digital Annealer (DA)~\cite{matsubara2020digitalannealer} in two complementary strategies. \textbf{Hybrid} supplies a DA-derived initial mapping to Qiskit's heuristic routing. \textbf{Full DA} solves mapping and routing as separate DA-assisted Quadratic Unconstrained Binary Optimization (QUBO) subproblems in an iterative workflow. Our primary contribution is a repeated, fixed-instance evaluation of DA-generated global placement followed by SABRE on a simplified 64-node $8\times8$ square-grid coupling graph; the evaluated Full DA workflow provides a complementary diagnostic of short-horizon routing quality and feasibility.

\section{Background and Related Work}
\label{sec:background}

\subsection{Annealing Machines and the Digital Annealer}

A QUBO minimizes \(E(\mathbf{x})=\mathbf{x}^{\top}Q\mathbf{x}\) over binary variables \(\mathbf{x}\in\{0,1\}^{N}\). We formulate the initial-mapping and routing subproblems as QUBOs, with objective terms representing interaction distance and qubit movement and penalty terms enforcing valid logical-to-physical assignments. Because validity is enforced through these penalties, their coefficients must be large enough to disfavor infeasible assignments while preserving objective-value differences among feasible solutions. Section~\ref{subsec:parameter_selection} reports the values fixed after calibration.

We use Fujitsu's fully connected Digital Annealer (DA)~\cite{matsubara2020digitalannealer}, which accepts dense mapping and routing QUBOs without minor embedding. This connectivity is relevant because the distance objectives and assignment penalties couple many pairs of binary variables. The DA service configuration used in this study supported QUBOs with up to \(8{,}192\) binary variables. This limit is specific to the available service configuration; other configurations may support larger problem sizes. Annealing is used as a heuristic, and we do not compare against an exact solver or report an optimality gap.

\subsection{Related QUBO-based Transpilation Methods}

QUBO-based qubit mapping is established prior work. ISAAQ~\cite{naito2023isaaqisingmachineassisted} partitions a circuit into local chunks and optimizes the mapping of each chunk independently using simulated annealing; adjacent mappings are then connected through heuristic routing. This decomposition reduces the QUBO size but cannot directly account for interactions spanning multiple chunks.

Our contribution is therefore not the first QUBO-based placement formulation. Hybrid instead optimizes a circuit-wide initial mapping before SABRE routing, accepting a larger QUBO in exchange for an objective that covers a broader set of circuit interactions. Full DA uses the same global-placement formulation with DA-assisted routing as a complementary diagnostic. The present fixed-instance evaluation compares output quality and feasibility under these configurations but does not measure scalability across circuit sizes.

\section{Proposed Method}
\label{chap:proposed_method}

\subsection{Target Device Topology}
\label{subsec:target_topology}

We use a simplified 64-node $8\times8$ square-grid coupling graph as the target topology.
Physical qubits use row-major indices $v=8r+c$, where $0\leq r,c<8$. Coupling edges join horizontal and vertical neighbors, with neither diagonals nor wraparound. In the routing constraints, $E$ is the symmetric directed arc set containing both orientations of every physical edge.

The topology is motivated by Fujitsu superconducting processors~\cite{FujitsuRiken2023, FujitsuCloud2023}. Square-lattice architectures are also used beyond Fujitsu, for example in IBM Nighthawk~\cite{IBMNighthawk2026}. Although our QUBO formulation depends only on the shortest-path distance matrix $d_{pq}$ and is therefore not inherently grid-specific, the present evaluation uses a simplified 64-node $8\times8$ square-grid coupling graph rather than an exact hardware coupling map. Consequently, the reported performance should not be assumed to transfer directly to other topologies such as heavy-hex, which remain future work.

\subsection{QUBO Formulation for Initial Mapping}
\label{subsec:qubo_formulation_of_mapping}

Both strategies share the same DA initial-mapping QUBO (Step 2 in Fig.~\ref{fig:Transpilation_flow}). It assigns logical qubits to physical qubits so that interacting pairs begin close together, reducing communication caused by SWAPs between non-adjacent locations. For the binary assignment variables $x_{i,p}$ in Table~\ref{tab:symbols}, the weighted distance objective is
\begin{equation}
H_{\mathrm{obj}} = \sum_{i,j} \sum_{p,q} w_{ij} \cdot d_{pq} \cdot x_{i,p} \cdot x_{j,q}
\end{equation}
where $x_{i,p}=1$ indicates that logical qubit $i$ is assigned to physical qubit $p$, $d_{pq}$ is the shortest-path distance between physical qubits, and $w_{ij}$ weights each interacting logical pair. The objective therefore favors assignments that place strongly weighted interactions close together.

For each logical pair $(i,j)$ that appears in at least one CNOT, $t_{ij}$ denotes the layer of its first interaction. We set $w_{ij}=w_{\mathrm{exp}}(t_{ij};0.10)$ using the exponential schedule defined in Eq.~\eqref{eq:exponential_decay}. This schedule assigns larger weights to earlier interactions while ensuring that later interactions continue to contribute to the objective. The selection of the schedule and its parameters is described in Section~\ref{subsec:parameter_selection}.

The assignment constraints are enforced by two penalty terms: $H_{\mathrm{const1}}$ assigns each logical qubit to exactly one physical qubit, and $H_{\mathrm{const2}}$ prevents two logical qubits from occupying the same physical qubit:
\begin{equation}
\begin{split}
H_{\mathrm{const}} &= H_{\mathrm{const1}} + H_{\mathrm{const2}} \\
&= \sum_i \left( \sum_p x_{i,p} - 1 \right)^2 + \sum_p \sum_{i \neq j} x_{i,p} x_{j,p}
\end{split}
\end{equation}

The first term is zero only when each logical qubit is assigned to exactly one physical qubit. The second term becomes positive whenever distinct logical qubits occupy the same physical qubit; unused physical qubits remain allowed when the logical circuit is smaller than the device.
The complete mapping QUBO is therefore
\begin{equation}
H_{\mathrm{map}} = H_{\mathrm{obj}} + \lambda_{\mathrm{map}} H_{\mathrm{const}}
\end{equation}
where $\lambda_{\mathrm{map}}$ controls the strength of the one-to-one mapping constraints; Section~\ref{subsec:parameter_selection} gives its value.
Minimizing $H_{\mathrm{map}}$ yields the initial mapping used by both Hybrid and Full DA by optimizing interactions across the entire input subcircuit rather than only the next executable gate.

\subsection{Proposed Method 1: Hybrid Strategy}
\label{subsec:hybrid_strategy}

Hybrid combines DA global search with high-speed classical routing. It first solves the subcircuit-wide mapping QUBO, which accounts for dependencies throughout the circuit, and passes the resulting physical layout to Qiskit for heuristic SWAP insertion and gate synthesis.

Specifically, Hybrid replaces Qiskit's layout stage with the DA-generated assignment while retaining the established SABRE routing procedure and its target-basis conversion. Since Qiskit and Hybrid otherwise use the same routing configuration, their comparison provides diagnostic evidence about the value of the DA starting point (Section~\ref{sec:evaluation}).

\begin{table}[t]
\caption{Definition of symbols used in the QUBO formulations.}
\label{tab:symbols}
\centering
\footnotesize
\setlength{\tabcolsep}{3pt}
\renewcommand{\arraystretch}{0.96}
\begin{tabularx}{\columnwidth}{@{}l>{\raggedright\arraybackslash}X@{}}
\toprule
Symbol & Description \\ \midrule
$i,j$ & Logical-qubit indices. \\
$p,q$ & Physical-qubit indices in $G(V,E)$. \\
$x_{i,p}$ & 1 iff logical $i$ is assigned to physical $p$. \\
$x_{i,p,t}$ & Mapping variable at routing step $t$. \\
$d_{pq}$ & Shortest-path distance between physical $p,q$. \\
$w_{ij}$ & Importance of the CNOT between logical $i,j$. \\
$\lambda_{\mathrm{map}},\lambda_{\mathrm{route}}$ & Mapping and routing constraint penalties. \\
$w_{\mathrm{swap}}$ & Qubit-movement cost weight. \\ \bottomrule
\end{tabularx}
\end{table}

\begin{table*}[!t]
\centering
\caption{Fixed transpilation configurations and repeated-trial protocols. Top-level counts do not imply equal internal search budgets.}
\label{tab:strategy_comparison}
\footnotesize
\setlength{\tabcolsep}{3.5pt}
\renewcommand{\arraystretch}{0.95}
\begin{tabularx}{\textwidth}{@{}l l l l >{\raggedright\arraybackslash}X l@{}}
\toprule
Method & Initial layout & Routing & Randomness & Top-level execution / internal work \\ \midrule
Hybrid & DA mapping & SABRE & DA + SABRE & 10 runs; one DA mapping and internal SABRE routing trials per run \\
Full DA & DA mapping & Iterative DA routing & DA sampling & 10 attempts; DA mapping and iterative DA routing per attempt \\
Qiskit & SABRE layout & SABRE & SABRE & 10 runs; one layout and four swap trials per run \\
t$\vert$ket$\rangle$ & Graph placement & LexiRoute & Deterministic & 10 executions; identical results in all runs \\ \bottomrule
\end{tabularx}
\end{table*}

\begin{table}[!t]
    \centering 
    \caption{\textbf{Characteristics of benchmark circuits.}} 
    \label{tab:circuit_comparison} 
    \begin{tabular}{lccl} \hline 
        Name & Qubits & CNOT count & Explanation \\ \hline
        BV & 30 & 18 & QASMBench~\cite{QASMBench_GitHub, Li2021QASMBench} \\
        Grover & 4 & 84 & Grover search \\
        QAOA-grid & 25 & 80 & Max-Cut (grid topology) \\
        GHZ & 50 & 49 & 50-qubit GHZ state \\
        QAOA-random & 30 & 105 & Max-Cut (3-regular graph) \\
        QV & 10 & 150 & Quantum Volume ($n=d$) \\
        ASP & 50 & 294 & Adiabatic state preparation \\
        \hline
    \end{tabular}
\end{table}

\subsection{Proposed Method 2: Full DA}
\label{subsec:full_da_strategy}

The Full DA strategy applies DA-based optimization to both the initial mapping (Section~\ref{subsec:qubo_formulation_of_mapping}) and the routing subproblem within an iterative workflow.

Because Full DA and Hybrid begin with the same mapping QUBO, their difference is the routing stage rather than the placement objective. Full DA explores whether extending global optimization to movement can further reduce gate overhead, while its bounded horizon and repeated solves expose the resource requirements of that extension.

\subsubsection{Transpilation Flow}
\label{subsec:Transpilation_flow}

\begin{figure}[!t] 
    \centering
    \includegraphics[width=0.82\columnwidth]{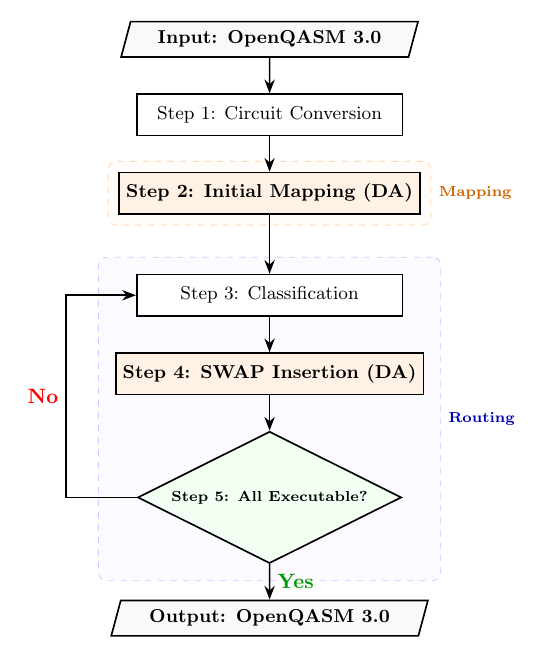}
    \caption{Full DA workflow: global initial mapping (Step 2) followed by iterative DA routing of non-executable gates (Steps 3--5).}
    \label{fig:Transpilation_flow}
\end{figure}

Full DA integrates DA-powered mapping and routing subproblems in the sequential flow of Fig.~\ref{fig:Transpilation_flow}. After circuit conversion (Step 1), it solves the global initial-mapping QUBO (Step 2). The transpiler then scans gates in execution order and classifies an interaction as executable when its mapped physical qubits are adjacent (Step 3). Non-executable gates are passed to the routing QUBO (Step 4); decoded movements update the mapping and introduce physical SWAP gates. On a successful attempt, classification and routing repeat until every gate satisfies the hardware couplings (Step 5), after which the hardware-compliant circuit is emitted.

Only the unresolved portion is presented to each routing solve. This iterative construction keeps each QUBO within the fixed temporal horizon, but unlike Hybrid it can require multiple remote DA calls before all circuit interactions become executable.

During iterative routing, a Full DA attempt was terminated as unsuccessful if either the number of executable CNOTs did not increase for ten consecutive routing trials, indicating stagnation, or a DA solve failed to return a decoded solution satisfying the assignment and movement constraints. Otherwise, classification and routing continued until all CNOTs were executable.

\subsubsection{QUBO Formulation for Routing}
\label{subsec:qubo_formulation_of_routing}
The routing phase searches for logical-qubit movements that make currently non-executable interactions adjacent. 
Variable $x_{i,p,t}$ (Table~\ref{tab:symbols}) indicates whether logical qubit $i$ is assigned to physical qubit $p$ at routing step $t$.
We fix the horizon at $T=2$ to respect the DA capacity: an instance with 64 logical and 64 physical qubits uses $64\times64\times2=8{,}192$ binary variables. Thus, each solve represents at most two coupling-edge moves per logical qubit; longer coordinated routes require successive calls, so short-term distance reductions need not yield an efficient global SWAP sequence.

The objective balances the communication distance of pending interactions against qubit movement:
\begin{equation}
H_{\mathrm{obj}} = H_{\mathrm{dist}} + w_{\mathrm{swap}} H_{\mathrm{swap}}
\end{equation}
where $w_{\mathrm{swap}}$ controls the relative movement contribution (Section~\ref{subsec:parameter_selection}).
For the non-executable part of the circuit, the distance term evaluates all weighted logical interactions at every routing step:
\begin{equation}
H_{\mathrm{dist}} = \sum_{t=1}^T \sum_{i,j} \sum_{p,q} w_{ij} \cdot d_{pq} \cdot x_{i,p,t} \cdot x_{j,q,t}
\end{equation}
Summing over $t$ keeps pending pairs close, with earlier dependencies weighted more strongly. When interactions share a logical qubit, this aggregate objective may favor a compromise placement rather than the path that most efficiently exposes the next gates.
The movement term detects changes in each assignment bit between consecutive routing steps, where $x_{i,p,0}$ denotes the pre-routing mapping:
\begin{equation}
H_{\mathrm{swap}} = \sum_{t=1}^{T} \sum_{i,p} ( x_{i,p,t-1} + x_{i,p,t} - 2 x_{i,p,t-1} x_{i,p,t})
\end{equation}
Each changed assignment bit contributes one unit of energy. Because a moved logical qubit changes multiple assignment bits and cyclic permutations require later decomposition, this term serves only as a proxy for SWAP count.

Physical validity requires a one-to-one logical--physical mapping at every $t$ and restricts each logical qubit to remain in place or traverse one coupling edge per step. 
Let $A=\{(p,p)\mid p\in V\}\cup E$ contain these allowed self-loops and device edges. 
The three terms enforce these constraints:
\begin{equation}
\begin{aligned}
H_{\mathrm{const}}={}&\sum_{t=1}^T\sum_i
 \left(\sum_p x_{i,p,t}-1\right)^2 \\
&+\sum_{t=1}^T\sum_p\sum_{i\ne j}x_{i,p,t}x_{j,p,t} \\
&+\sum_{t=1}^T\sum_i\sum_{(p,q)\notin A}
 x_{i,p,t-1}x_{i,q,t}.
\end{aligned}
\end{equation}
The last term assigns energy to every transition outside $A$, prohibiting non-adjacent movement while allowing a qubit to stay in place. The complete routing QUBO is
\begin{equation}    
H_{\mathrm{route}} = H_{\mathrm{obj}} + \lambda_{\mathrm{route}} H_{\mathrm{const}}
\end{equation}
where $\lambda_{\mathrm{route}}$ controls routing-constraint strength (Section~III-E).
A feasible decoded solution defines topology-valid movement paths.

\subsubsection{Decoding and Post-processing for Routing}
\label{subsubsec:decoding_post_processing}

The decoder compares mappings at $t-1$ and $t$ and converts their changes into a sequence of physical SWAPs. A solution may contain a simultaneous closed permutation, or ``cyclic swap'' (e.g., $q_1\to p_2,q_2\to p_3,q_3\to p_1$). Such a permutation is logically valid, but prohibiting every cycle inside the QUBO would require excessive auxiliary variables. Classical post-processing therefore detects each cycle and decomposes it into pairwise SWAPs along device edges. This maintains the logical state and topology constraints without enlarging the QUBO.

\subsection{QUBO Parameter Selection}
\label{subsec:parameter_selection}

To select the interaction-weight schedule, we compared the linear, exponential, and polynomial decay functions defined in Eqs.~\eqref{eq:linear_decay}--\eqref{eq:polynomial_decay}, respectively.

The linear decay function was defined as
\begin{equation}
w_{\mathrm{lin}}(t;p)
=
\max\left\{
1,\,
W_{\max}
-
\frac{t(W_{\max}-1)}{pT_{\max}}
\right\},
\label{eq:linear_decay}
\end{equation}
where $p\in\{0.1,0.3,0.5,1.0\}$.

The exponential decay function was defined as
\begin{equation}
w_{\mathrm{exp}}(t;k)
=
W_{\max}\exp(-kt)+1,
\label{eq:exponential_decay}
\end{equation}
where $k\in\{0.01,0.05,0.10,0.50\}$.

The polynomial decay function was defined as
\begin{equation}
w_{\mathrm{poly}}(t;\alpha)
=
W_{\max}
\left(
1-\frac{t}{T_{\max}}
\right)^{\alpha},
\label{eq:polynomial_decay}
\end{equation}
where $\alpha\in\{1.5,2,3\}$.

We compared the schedules in Eqs.~\eqref{eq:linear_decay}--\eqref{eq:polynomial_decay} by transpiling randomly generated calibration circuits for a 64-node $8\times8$ square-lattice topology. 
We evaluated each candidate schedule using five randomly generated calibration circuits. For each schedule, we measured (i) the mean number of CNOTs that were executable without SWAP insertion under the initial mapping and (ii) the mean total coupling-graph distance between the physical qubits assigned to the operands of all CNOTs. The exponential schedule in Eq.~\eqref{eq:exponential_decay}, with $k=0.10$, achieved both the largest mean number of directly executable CNOTs and the smallest mean total distance. We therefore adopted this schedule for all reported experiments. The calibration circuits were disjoint from the evaluation benchmarks.
Mapping used $W_{\max}=1000$ and $\lambda_{\mathrm{map}}=10\sum_t w(t)$. Routing used $w_{\mathrm{swap}}=0.009$ and $\lambda_{\mathrm{route}}=5\sum_t w(t)$. The routing horizon remained $T=2$ because of the variable-capacity limit described in Section~\ref{subsec:qubo_formulation_of_routing}.

\begin{table*}[!t]
    \centering
    \caption{
        Equivalent CNOT counts for the five transpilation methods. \\
        Values are the mean $\pm$ sample standard deviation over repeated trials. \\
        Red values indicate the lowest mean equivalent CNOT count for each circuit.
    }
    \label{tab:all_method_comparison}
    \setlength{\tabcolsep}{4pt}
    \renewcommand{\arraystretch}{1.05}

    \begin{tabular}{lccccccc}
        \toprule
        \textbf{Method}
        & \textbf{BV}
        & \textbf{GHZ}
        & \textbf{QAOA-grid}
        & \textbf{Grover}
        & \textbf{QAOA-random}
        & \textbf{QV}
        & \textbf{ASP} \\
        \midrule

        Hybrid
        & \textcolor{red}{$33.5 \pm 3.2$}
        & \textcolor{red}{$49.0 \pm 0.0$}
        & \textcolor{red}{$80.0 \pm 0.0$}
        & \textcolor{red}{$112.2 \pm 1.0$}
        & \textcolor{red}{$150.9 \pm 4.7$}
        & \textcolor{red}{$196.8 \pm 4.3$}
        & \textcolor{red}{$245.0 \pm 0.0$}
        \\

        Qiskit
        & $46.2 \pm 6.6$
        & \textcolor{red}{$49.0 \pm 0.0$}
        & \textcolor{red}{$80.0 \pm 0.0$}
        & $112.6 \pm 0.8$
        & $169.4 \pm 14.2$
        & $205.5 \pm 6.7$
        & \textcolor{red}{$245.0 \pm 0.0$}
        \\

        t$\vert$ket$\rangle$
        & $60.0 \pm 0.0$
        & \textcolor{red}{$49.0 \pm 0.0$}
        & \textcolor{red}{$80.0 \pm 0.0$}
        & $156.0 \pm 0.0$
        & $255.0 \pm 0.0$
        & $225.0 \pm 0.0$
        & \textcolor{red}{$245.0 \pm 0.0$}
        \\

        Full DA
        & N/F
        & \textcolor{red}{$49.0 \pm 0.0$}
        & \textcolor{red}{$80.0 \pm 0.0$}
        & $201.3 \pm 10.8$
        & $469.9 \pm 130.2$
        & $832.6 \pm 38.0$
        & $294.0 \pm 0.0$
        \\

        ISAAQ
        & $119.4 \pm 13.9$
        & $375.8 \pm 62.6$
        & $323.5 \pm 37.6$
        & $246.6 \pm 19.7$
        & $641.6 \pm 60.8$
        & $795.7 \pm 59.8$
        & $3251.6 \pm 92.5$
        \\

        \bottomrule
        \multicolumn{8}{l}{\footnotesize N/F: no feasible output.} \\
        \multicolumn{8}{l}{\footnotesize Full DA feasible rate: BV 0/10, QV 8/10 (statistics computed from the eight feasible attempts).}
    \end{tabular}
\end{table*}

\begin{figure*}[!t]
    \centering

    \begin{minipage}[t]{0.48\textwidth}
        \centering
        \includegraphics[width=\linewidth]{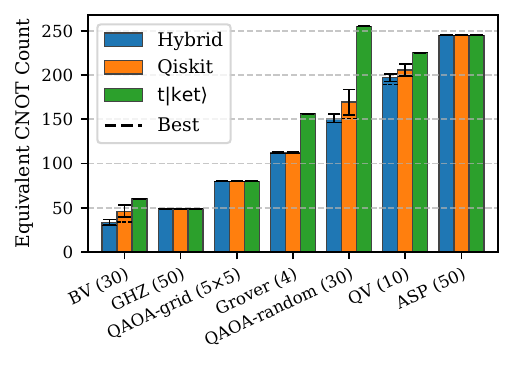}
        \caption{\textbf{Equivalent CNOT counts over ten independent executions of Hybrid, Qiskit, and t$\vert$ket$\rangle$ for each benchmark.}
        Bars show means, error bars show sample standard deviations, and dashed markers show the observed minima.}
        \label{fig:hybrid_cnot_comparison}
    \end{minipage}
    \hfill
    \begin{minipage}[t]{0.48\textwidth}
        \centering
        \includegraphics[width=\linewidth]{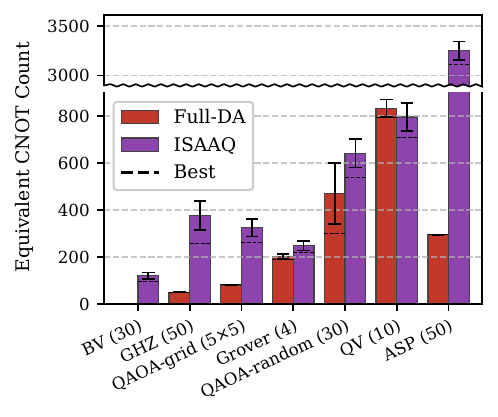}
        \caption{\textbf{Equivalent CNOT counts from ten Full DA attempts and ten ISAAQ executions per benchmark.}
        Full DA statistics are computed over feasible outputs only.
        Bars show means, error bars show sample standard deviations, and dashed markers show the observed minima.}
        \label{fig:fullda_cnot_comparison}
    \end{minipage}

\end{figure*}
\section{Evaluation}
\label{sec:evaluation}

We report three comparison tracks: Hybrid versus the configured Qiskit and t$\vert$ket$\rangle$ baselines; Full DA versus ISAAQ~\cite{naito2023isaaqisingmachineassisted}; and a diagnostic comparison of Hybrid and Full DA where Full DA produced feasible outputs. BV is included only in the Full DA feasibility assessment because none of its ten attempts was feasible. These tracks are evaluated separately and do not constitute an all-method ranking.

\subsection{Experimental Setup}
\label{subsec:experimental_setup}

Experiments used Python 3.10, Qiskit 2.3.1, pytket 2.13.0, Amplify 1.4.1~\cite{fixstars_amplify}, and Fujitsu Digital Annealer Gen4. All methods targeted the same 64-node $8\times8$ grid defined in Section~\ref{subsec:target_topology}. Equivalent CNOT count assigns one count to each CNOT and three to each explicit SWAP. This structural metric does not include hardware error rates, decoherence, or execution data and therefore does not directly measure fidelity.

Table~\ref{tab:strategy_comparison} summarizes the method configurations and repeated-trial protocols. Qiskit used optimization level 3, the basis \texttt{["u","cx"]}, one layout trial, four swap trials, four optimization iterations, and seeds \(10,20,\ldots,100\). Hybrid supplied a DA-generated initial mapping while retaining the same basis and SABRE routing settings. For t$\vert$ket$\rangle$, \texttt{GraphPlacement} was followed by \texttt{LexiLabellingMethod} and \texttt{LexiRouteRoutingMethod}; Full DA used the iterative workflow in Fig.~\ref{fig:Transpilation_flow}. Internal solver trials and samples were not treated as independent observations, and equal top-level execution counts do not imply matched computational budgets.

Each method--benchmark configuration was executed or attempted ten times. For each pair, \(N\) denotes the successful top-level executions used to compute the arithmetic mean, sample SD with \(N-1\) normalization, range, and observed minimum. A Full DA attempt was considered feasible when it returned a routed circuit with a recorded equivalent-CNOT count. For benchmark \(c\), the Full DA reduction relative to ISAAQ is \(r_c^{\mathrm{FD}}=100(I_c-F_c)/I_c\), and the reported summary is the unweighted arithmetic macro-average over benchmarks with at least one feasible Full DA output. It therefore characterizes feasible-output quality, not feasibility or computational cost.

\subsection{Benchmark Circuit Generation}
\label{subsec:benchmark_circuit_generation}

Table~\ref{tab:circuit_comparison} summarizes seven fixed benchmark instances spanning concentrated, structured, and random interaction patterns. Each family is represented by one instance, so the evaluation does not measure between-instance variability. Bernstein--Vazirani (BV), taken from QASMBench~\cite{QASMBench_GitHub, Li2021QASMBench}, has CNOT gates sharing a common target qubit. Greenberger--Horne--Zeilinger (GHZ) applies a Hadamard gate followed by a linear CNOT chain. Grover uses an oracle marking \(\ket{11\dots1}\) and the theoretical iteration count \(\lfloor\frac{\pi}{4}\sqrt{2^n}\rfloor\).
The two \(p=1\) quantum approximate optimization algorithm (QAOA) benchmarks differ in interaction structure: QAOA-grid uses a 2D grid matching the target coupling graph, whereas QAOA-random uses a NetworkX-generated random 3-regular graph with seed 42. Quantum Volume (QV) was generated with Qiskit using \(n=d=10\) and seed 42. Adiabatic state preparation (ASP) uses three Trotter steps, each comprising single-qubit \(X\) rotations and nearest-neighbor \(ZZ\) interactions along a one-dimensional chain.

Fixed seeds keep the QAOA-random and QV instances identical across methods. Repeated executions therefore characterize stochastic variation in transpilation rather than variability among benchmark instances.

\subsection{Performance Evaluation}
\label{subsec:performance_evaluation}

Figure~\ref{fig:hybrid_cnot_comparison} compares Hybrid with the configured Qiskit and t$\vert$ket$\rangle$ baselines. Hybrid achieved lower mean equivalent CNOT counts than both configured baselines on four of the seven benchmarks and identical means on the remaining three. The unweighted arithmetic macro-average over all seven benchmark-level reductions was 6.14\% relative to Qiskit and 17.9\% relative to t$\vert$ket$\rangle$, with maximum reductions of 27.5\% and 44.2\%, respectively. All percentage reductions are computed from the corresponding top-level means; dashed minima indicate the best observed results.

Table~\ref{tab:all_method_comparison} shows that the three ties occurred on GHZ, QAOA-grid, and ASP, whose chain- or grid-structured interactions can be embedded naturally in the square-lattice topology. In particular, the equivalent CNOT counts for GHZ and QAOA-grid were equal to their input CNOT counts, indicating that the configured methods introduced no additional two-qubit-gate overhead and leaving little room for improvement through initial placement. By contrast, BV contains interactions concentrated on a common target, while QAOA-random and QV contain less hardware-aligned interaction patterns. The improvements on these benchmarks are therefore consistent with DA-generated global placement being most useful when the circuit is sensitive to its initial layout. The reduction on Grover was comparatively small relative to the observed run-to-run variation and should not be interpreted as evidence of a robust advantage by itself. Because each benchmark family is represented by only one fixed instance, these structure-dependent interpretations remain qualitative rather than general conclusions about the circuit families.

Hybrid uses DA-generated initial placement followed by the same SABRE routing strategy as the configured Qiskit baseline; its lower counts therefore provide descriptive evidence that the DA-generated starting layout can benefit subsequent heuristic routing. In contrast, among the six benchmarks with feasible Full DA outputs, Full DA had higher means than Hybrid on four and tied on two, with no lower result. The ties on GHZ and QAOA-grid are consistent with their limited routing requirements, whereas the degradation on the other benchmarks is consistent with the evaluated \(T=2\) routing horizon favoring short-term distance reductions that do not necessarily form an efficient global SWAP sequence.

Figure~\ref{fig:fullda_cnot_comparison} compares feasible Full DA outputs with ISAAQ. Full DA achieved lower means on five of the six comparable benchmarks: the unweighted arithmetic macro-average reduction was 48.9\%, including the 4.6\% increase on QV, and the maximum reduction was 91.0\% on ASP. These results do not by themselves establish that DA-assisted routing is superior, because the comparison reflects the combined effects of the evaluated placement, routing, and solver configurations, and Full DA did not outperform Hybrid on any benchmark with feasible outputs. Full DA also produced feasible outputs in 0/10 BV attempts, 8/10 QV attempts, and 10/10 attempts for each remaining benchmark. The reported reductions therefore characterize success-conditioned output quality and should be interpreted together with feasibility rather than as unconditional performance. Taken together, the results identify DA-generated global placement followed by heuristic routing as the primary positive result, while the Full DA results characterize the quality and feasibility limitations of the evaluated short-horizon routing workflow.

\subsection{Execution Time Analysis}
\label{subsec:execution_time}

Table~\ref{tab:runtime_comparison} summarizes the runtime characteristics of the evaluated methods. Because Hybrid and Full DA invoke the DA a different number of times, their runtimes should not be interpreted as a controlled speedup comparison. Communication accounted for more than half of the runtime in both DA-based methods; because this latency depends on network and service conditions, the reported values are indicative, and whether a co-located environment such as ABCI-Q~\cite{abci_q_website} can reduce this overhead remains to be evaluated under controlled conditions.

\begin{table}[t]
\centering
\caption{Compilation runtime and DA communication overhead \\ across six benchmarks, reported as mean [minimum--maximum] of benchmark-level means.}
\label{tab:runtime_comparison}
\begin{tabular}{lcc}
\toprule
Method & Runtime (s) & DA communication overhead (\%) \\
\midrule
Full DA & 351.72 [23.30--956.65] & 56.0 [55.2--57.4] \\
Hybrid & 25.95 [22.77--39.89] & 58.2 [54.9--73.8] \\
ISAAQ & 5.83 [5.10--7.42] & -- \\
t$\vert$ket$\rangle$ & 1.183 [0.391--1.945] & -- \\
Qiskit & 0.0220 [0.0089--0.0560] & -- \\
\bottomrule
\end{tabular}
\end{table}
\section{Conclusion}
\label{sec:conclusion}

We evaluated DA-generated global placement for CNOT-count-oriented transpilation on seven fixed standalone circuit instances. Under the evaluated configurations, Hybrid achieved lower mean equivalent CNOT counts than both configured baselines on four instances and identical means on the remaining three, suggesting that global initial placement can benefit subsequent heuristic routing on placement-sensitive circuits.

Full DA achieved lower means than ISAAQ on five of the six benchmarks with feasible outputs, but it did not outperform Hybrid and produced no feasible output for BV. These results characterize success-conditioned output quality while also exposing limitations in the feasibility and global routing quality of the evaluated \(T=2\) workflow. Because the evaluation uses one fixed instance per benchmark family and equivalent CNOT count as a structural proxy, it does not establish scalability across circuit families, an end-to-end partitioning-runtime tradeoff, or hardware-fidelity improvement.

Future work will distinguish fixed-horizon representation limits from QUBO-optimization limits, incorporate hardware noise, evaluate additional coupling topologies, and extend the benchmark set toward complete circuit-partitioning workflows.

\section*{Acknowledgment}

This paper is based on results obtained from a project,
JPNP23003, commissioned by the New Energy and Industrial Technology Development Organization (NEDO).
KW and TS are also supported by JST COI-NEXT Grant Number JPMJPF2221.
SN is also supported by the JSPS Overseas Research Fellowship.

\clearpage
\balance

\bibliographystyle{IEEEtran}
\bibliography{main} 
\vspace{12pt}

\end{document}